\documentclass{emulateapj}

\usepackage{lscape}

\slugcomment{Accepted for publication in the Astrophysical Journal: August 20, 2008}

\shorttitle{The nature of DOGs}
\shortauthors{Pope et al.}

\begin{document}

\title{The nature of faint {\it Spitzer}-selected dust-obscured galaxies}

\author{Alexandra Pope\altaffilmark{1,2,$\ast$},
R.~Shane Bussmann\altaffilmark{3},
Arjun Dey\altaffilmark{2},
Nicole Meger\altaffilmark{4},
David M.~Alexander\altaffilmark{5},
Mark Brodwin\altaffilmark{2}, 
Ranga-Ram Chary\altaffilmark{6},
Mark E.~Dickinson\altaffilmark{2}, 
David T.~Frayer\altaffilmark{7},
Thomas R.~Greve\altaffilmark{8},
Minh Huynh\altaffilmark{7}, 
Lihwai Lin\altaffilmark{9},
Glenn Morrison\altaffilmark{10, 11}, 
Douglas Scott\altaffilmark{4},
Chi-Hung Yan\altaffilmark{9, 12}
}

\altaffiltext{$\ast$}{Based on observations obtained at the Canada-France-Hawaii Telescope (CFHT) which is operated by the National Research Council of Canada, the Institut National des Sciences de l'Univers of the Centre National de la Recherche Scientifique of France, and the University of Hawaii.}
\altaffiltext{1}{{\it Spitzer} Fellow; pope@noao.edu}
\altaffiltext{2}{National Optical Astronomy Observatory, 950 N. Cherry Ave., Tucson, AZ, 85719}
\altaffiltext{3}{Steward Observatory, Department of Astronomy, University of Arizona, 933 N.~Cherry Ave., Tucson, AZ 85721}
\altaffiltext{4}{Department of Physics \& Astronomy, University of British Columbia, Vancouver, BC, V6T 1Z1, Canada}
\altaffiltext{5}{Department of Physics, Durham University, Durham, DH1 3LE, UK}
\altaffiltext{6}{{\em Spitzer} Science Center, MS 220-6 Caltech, Pasadena, CA 91125}
\altaffiltext{7}{Infrared Processing and Analysis Center, California Institute of Technology 100-22, Pasadena, CA 91125 USA}
\altaffiltext{8}{Max-Planck Institute f\"{u}r Astronomie, K\"{o}nigstuhl 17, Heidelberg, D-69117, Germany}
\altaffiltext{9}{Institute of Astronomy \& Astrophysics, Academia Sinica, Taipei 106, Taiwan}
\altaffiltext{10}{Institute for Astronomy, University of Hawaii, Honolulu, HI, 96822, USA}
\altaffiltext{11}{Canada-France-Hawaii Telescope, Kamuela, HI, 96743, USA}
\altaffiltext{12}{Department of Earth Sciences, National Taiwan Normal University}

\begin{abstract}
We use deep far-IR, submm, radio and X-ray imaging and mid-IR spectroscopy to explore the nature of a sample of {\it Spitzer}-selected dust-obscured galaxies (DOGs) in GOODS-N. A sample of 79 galaxies satisfy the criteria $R-[24]>14$ (Vega) down to $S_{24}>100\,\mu$Jy (median flux density $S_{24}=180\,\mu$Jy). Twelve of these galaxies have IRS spectra available which we use to measure redshifts and classify these objects as being dominated by star formation or active galactic nuclei (AGN) activity in the mid-IR. The IRS spectra and {\it Spitzer} photometric redshifts confirm that the DOGs lie in a tight redshift distribution around $z\sim2$. 
Based on mid-IR colors, 80\% of DOGs are likely dominated by star formation; the stacked X-ray emission from this sub-sample of DOGs is also consistent with star formation. 
Since only a small number of DOGs are individually detected at far-IR and submm wavelengths, we use a stacking analysis to determine the average flux from these objects and plot a composite IR (8--1000$\,\mu$m) spectral energy distribution (SED). 
The average luminosity of these star forming DOGs is $L_{\rm{IR}}\sim1\times10^{12}L_{\odot}$. We compare the average star forming DOG to the average bright ($S_{850}>5\,$mJy) submillimeter galaxy (SMG); the $S_{24}>100\,\mu$Jy DOGs are 3 times more numerous but 8 times less luminous in the IR. The far-IR SED shape of DOGs is similar to that of SMGs (average dust temperature of around 30$\,$K) but DOGs have a higher mid-IR to far-IR flux ratio. 
The average star formation-dominated DOG has a star formation rate of $200\,\rm{M_{\odot}\,yr^{-1}}$ which, given their space density, amounts to a contribution of $0.01\,\rm{M_{\odot}\,yr^{-1}Mpc^{-3}}$ (or 5--10\%) to the star formation rate density at $z\sim2$. 
We use the composite SED to predict the average flux of DOGs in future {\it Herschel}/PACS $100\,\mu$m and SCUBA-2 $450\,\mu$m surveys and show that the majority of them will be detected. 
\end{abstract}


\section{Introduction}
\label{sec:intro}

Large extragalactic surveys with the {\it Spitzer Space Telescope} (Werner et al.~2004) have revealed many high redshift objects which are bright in the mid-infrared (mid-IR) and have red mid-IR to optical colors (e.g.~Houck et al.~2005; Yan et al.~2007). The selection in the mid-IR indicates that these objects are dusty; however, without mid-IR spectra it is not clear whether the dust is heated by active galactic nuclei (AGN) or star formation activity or both. Furthermore, extrapolating from mid-IR to total IR luminosity is uncertain without good constraints spanning the far-IR dust peak (e.g.~Papovich et al.~2007; Daddi et al.~2007). 

A sample of IR-luminous Dust Obscured Galaxies (DOGs), selected to have very red $R$-[24] color, was recently presented by Dey et al.~(2008, hereafter D08). From spectroscopic observations DOGs were found to have a tight redshift distribution around $z\sim2$, very similar to that of the submillimeter selected galaxies (SMGs, e.g. Chapman et al.~2005). The space density and clustering of DOGs are also comparable to those of the bright ($S_{850}>6\,$mJy) SMGs, suggesting that these two populations might be associated (e.g., in an evolutionary sequence; D08, Brodwin et al.~2008). 

The origin of the bolometric luminosity in DOGs is uncertain. Using a similar $R$--[24] selection and an additional $R$--$K$ color criterion, Fiore et al. (2008, hereafter F08) selected a sample of faint DOGs ($S_{24}>40\,\mu$Jy) in the CDF-S and concluded, based on their stacked X-ray spectrum, that 80\% are Compton-thick AGN. 

In order to test the relationship between SMGs, DOGs, and the F08 Compton-thick AGN, one needs multi-wavelength data including near-IR, far-IR and submillimeter observations of a large sample of DOGs. 
In this paper we use the deep multi-wavelength data in the GOODS-N field (Giavalisco et al.~2004) to study a sample of faint DOGs in order to put constraints on their infrared luminosities, determine the role of AGN and star formation activity in these systems and compare them with SMGs. Our goal is to improve our understanding of the role of DOGs in massive galaxy evolution. 

All magnitudes in this paper use the Vega system unless otherwise noted. We assume a standard cosmology with $H_{0}=72\,\rm{km}\,\rm{s}^{-1}\,\rm{Mpc}^{-1}$, $\Omega_{\rm{M}}=0.3$ and $\Omega_{\Lambda}=0.7$. 

\section{Data and sample selection}
\label{sec:data}

Using the Subaru $R$-band (Capak et al.~2004) and {\it Spitzer} 24$\,\mu$m (Chary et al.~in preparation) data in GOODS-N we select 79 galaxies which satisfy the DOGs criteria of $R$--[24]$>14$ (i.e.~$S_{24}/S_{R}\ge1000$) and $S_{24}>100\,\mu$Jy. The 24$\,\mu$m fluxes are measured by fitting the point spread function using the IRAC positions as priors (see Chary et al.~in preparation for details), and the $R$-band photometry is measured in 3 arcsecond diameter apertures since this gave optimal SNRs. This GOODS-N sample is significantly deeper than the Bootes DOGs sample of D08, which was limited by $S_{24}>300\,\mu$Jy; there are only 13 DOGs in GOODS-N above this flux cut. The median $24\,\mu$m flux density for the 79 DOGs in GOODS-N is 180$\,\mu$Jy. 

To understand the nature of DOGs we exploit the deep multi-wavelength data available in GOODS-N. 
We measure the $K$-band magnitude of each DOG in the new WIRCAM/CFHT $K$-band images (Lin et al.~in preparation), and the $B$ and $z$ magnitudes from the Capak et al.~(2004) Subaru images in matched circular apertures. 
For comparison we also measure the $R$ and $K$ magnitudes for the sample of SMGs in GOODS-N (Pope et al.~2006). 
The IRAC fluxes of all DOGs are available from the deep {\it Spitzer} legacy images (Dickinson et al.~in preparation), where we use 4 arcsecond diameter apertures (with the appropriate aperture corrections applied) for photometry. 

Many DOGs are not individually detected in the X-ray, far-IR, (sub)mm, and radio (see Table \ref{tab:ids}) and so we must rely on stacking analyses. We use the {\it Chandra X-ray Observatory} 2 Msec image (Alexander et al.~2003) to identify DOGs which are formally detected and also to perform a stacking analysis of the undetected sources.
MIPS 70 and 160$\,\mu$m data were reduced following the techniques of Frayer et al.~(2006a,b). The MIPS 70 and 160$\,\mu$m images are searched for detections as well as used for stacking analysis (see Huynh et al.~2007 for details). 
In the (sub)mm, we use the 850$\,\mu$m Submillimetre Common User Bolometer Array (SCUBA) super-map (Borys et al.~2003; Pope et al.~2005) and the 1.2$\,$mm Max-Planck Millimetre Bolometer (MAMBO) map (Greve et al.~2008). 
For (sub)mm stacking we perform a variance-weighted average due to the variable noise levels across the maps and stack on an image which has the point sources removed so as not to bias the result. 
At the longest wavelengths, we use the 1.4$\,$GHz VLA A+B-array map and catalogs of Morrison et al.~(in preparation). 
Stacking in the radio and far-IR is performed by stacking images centered on each source and therefore we can check to make sure that the resulting point spread function is as expected. 
For the stacking analyses in the X-ray, MIPS, SCUBA, MAMBO and radio images we perform Monte Carlo simulations at random positions to determine the error and significance of the stacked result.

Several {\it Spitzer} IRS spectroscopy programs have targeted GOODS-N; we find twelve DOGs have spectra available from GO-20456 (10 sources, Pope et al.~2008; Murphy et al.~in preparation) and GO-20733 (2 sources, J.~Van Duyne private communication). Details of the IRS data reduction can be found in Pope et al.~(2008). 

\section{Results}

\subsection{Comparison with other selection criteria}

\begin{figure}
\plotone{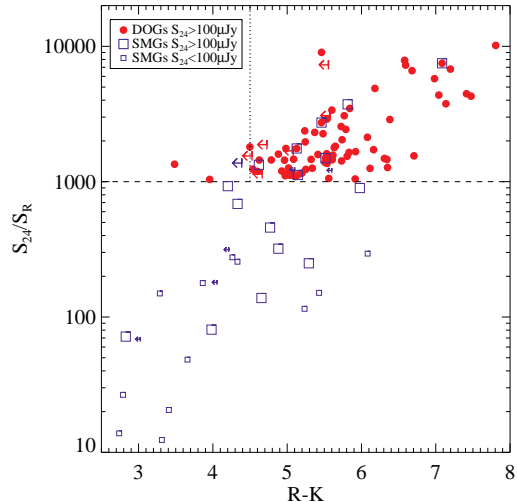}
\caption{$S_{24}$/$S_{R}$ as a function of $R$--$K$ color for DOGs and SMGs in GOODS-N. The dashed line indicates the DOGs selection criteria while the dotted line shows the additional color constraint in Fiore et al.~(2008). At least 90\% of DOGs and 20\% of SMGs meet the Fiore et al.~(2008) criteria (30\% of SMGs are classified as DOGs). 
\label{fig:fiore}}
\end{figure}

Fig.~\ref{fig:fiore} shows the $R$--$K$--$S_{24}$ color-color plot (cf., F08) for the DOGs and the GOODS-N SMGs (Pope et al.~2006). While SMGs are not constrained to any region of the diagram, the DOGs almost all satisfy the additional $R-K$ criterion from F08 (dotted line; F08 conclude that 80\% of galaxies in this region are Compton-thick AGN). This is somewhat unexpected, since Fig.~3 of F08 shows just as many $S_{24}/S_{R}\ge1000$ galaxies on either side of $R-K=4.5$. However F08 push even deeper ($S_{24}>40\,\mu$Jy) and it is clear that the fainter $24\,\mu$m samples contain a higher fraction of bluer (in $R-K$) objects (F.~Fiore, private communication). 
Fig.~\ref{fig:fiore} also shows that 30\% ($>20$\%) of SMGs meet the DOGs (F08) criteria, respectively. It is difficult to assess the fraction of DOGs which are submm detected, due to the highly varying noise levels in the SCUBA map of GOODS-N (Borys et al.~2003), but looking only at the low noise regions ($<2.3\,$mJy RMS) of the map it appears that around 30\% ($7/24$) of DOGs are formally detected ($>3.5\sigma$) at 850$\,\mu$m. In contrast, only $2/73$ DOGs are coincident with detections in the GOODS-N MAMBO map, down to a 1.2$\,$mm depth of $\sim0.8\,$mJy RMS (Greve et al.~2008). The difference between the submm and mm detection rates of DOGs could be because the millimeter selection picks out either higher redshift or cooler objects than the submillimeter (see discussion in Greve et al.~2008). 

\begin{figure}
\plotone{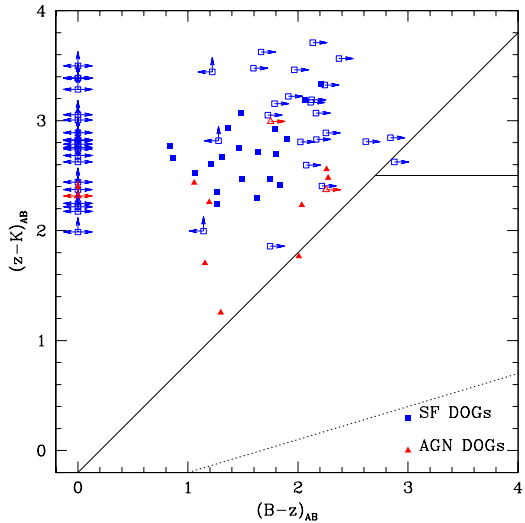}
\caption{{\it BzK} color-color plot for DOGs in GOODS-N. The diagonal line separates active galaxies at $z=1.4$--2.5 (above line, Daddi et al.~2004). Squares and triangle are the GOODS-N DOGs where the different symbols show the mid-IR SF and AGN classified DOGs (see Section \ref{sec:agnsf}). 
\label{fig:bzk}}
\end{figure}

Another widely-used selection criteria for high redshift galaxies is the {\it BzK} color-color plot (Daddi et al.~2004). Fig.~\ref{fig:bzk} shows the {\it BzK} plot for the DOGs in GOODS-N. All DOGs with a $>3\sigma$ detection in $K$ are plotted ($75/79$); many of these are only limits at $B$ and $z$.  While almost all DOGs are consistent with the {\it BzK} selection, many would not be included in robust active {\it BzK} samples since they are not detected in $B$ and/or $z$. 
Roughly 12\% of {\it BzK} galaxies down to $S_{24}>100\,\mu$Jy will satisfy the DOGs criteria. 
The overlap between DOGs and {\it BzK} galaxies is an interesting topic, and will be explored in more detail in a future paper (Meger et al.~in preparation). 
The position of the DOGs within the {\it BzK} plot is our first clue that these are star-forming galaxies at $z=1.4$--2.5 (Daddi et al.~2004); the redshifts, star formation and AGN activity of DOGs will be discussed in the following sections.

\subsection{Redshifts}

A redshift distribution peaking around 2 is expected for the DOGs selection criteria since the 24$\,\mu$m detection will limit the sample to sources below $z\simeq3$ and the red mid-IR to optical color allows only for sources which are very faint in the optical, which weeds out many $z<1$ sources. 
D08 presented a redshift distribution for bright ($S_{24}>0.3\,$mJy) DOGs based on spectroscopically measured redshifts, resulting in a Gaussian distribution with \={z}=2.0 and $\sigma(z)$=0.5. D08 also note no obvious dependence of redshift on 24$\,\mu$m flux within their sample. 
F08 presented a photometric redshift distribution for galaxies which meet their two color selections down to $S_{24}>40\,\mu$Jy, and it is remarkably similar to that of the D08 DOGs. This suggests that our $S_{24}>100\,\mu$Jy sample of DOGs will also have a similar redshift distribution to that of the brighter DOGs in D08. 

The redshifts for the 12 GOODS-N DOGs with IRS spectra range from 1.6--2.6 with a peak at 2, consistent with the spectroscopic redshift distribution from D08. However, 10/12 of the GOODS-N DOGs with IRS spectra are brighter than $S_{24}>300\,\mu$Jy. In addition to the 12 DOGs with IRS spectra, only 1 additional DOG has an optical spectroscopic redshift from the $\sim3000$ spectroscopic redshifts available in GOODS-N emphasizing how faint these galaxies are in the optical. This DOG is the unusual Waddington et al.~(1999) dusty radio galaxy at $z=4.424$; the highest confirmed redshift $24\,\mu$m source in GOODS-N.

\begin{figure}
\plotone{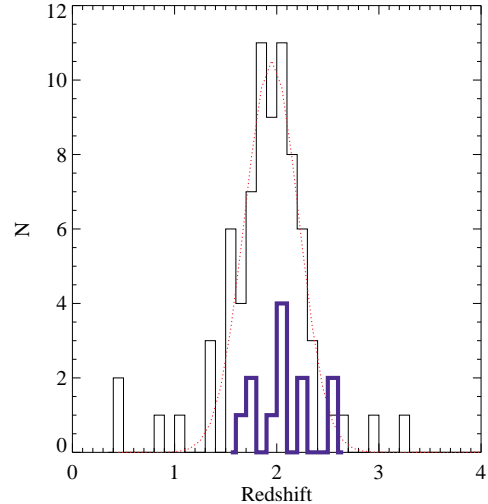}
\caption{Redshift distribution for GOODS-N DOGs using the Pope et al.~(2006) photometric redshift estimator (thin solid distribution). The thick solid histogram shows the DOGs with IRS spectroscopic redshifts. The dotted curve is the Gaussian fit to the photometric redshifts which gives $<z>=2.0\pm0.3$.
\label{fig:zdist}}
\end{figure}

Although most DOGs are too faint in the optical to yield accurate traditional optical photometric redshifts, we can use the IRAC photometry as a rough estimate of the redshift since at $z=2$ these channels sample the 1.6$\,\mu$m stellar bump (e.g.~Simpson \& Eisenhardt 1999, Sawicki 2002). 
Recently, Farrah et al.~(2008) presented a study of `bump2' sources defined as $S_{3.6}<S_{4.5}>S_{5.8}$ and $S_{4.5}>S_{8.0}$ and found them to lie in a tight redshift distribution, $<z>=1.71\pm0.15$. 
Similarly, we can select `bump3' sources ($S_{4.5}<S_{5.8}>S_{8.0}$ and $S_{3.6}<S_{5.8}$) which should lie around $z\simeq2.5$. 
Of the 79 DOGs in GOODS-N, 62 (78\%) satisfy either the `bump2' or `bump3' criteria which places them between $z=1.3$--$2.9$; this is consistent with the spectroscopic redshift distribution of DOGs in D08. 
The presence of a stellar bump in most DOGs also indicates that they are not dominated by AGN emission in the rest-frame near-IR; this will be discussed further in the next section. 

As an additional test of the redshifts of DOGs, we used the independent photometric-redshift estimate derived in Pope et al.~(2006). 
Equation 2 of Pope et al.~(2006) provides a simple empirical relation between the redshift and the IRAC and MIPS 24$\,\mu$m photometry which has been tuned to SMGs with spectroscopic redshifts. 
The application of this relation to the DOGs may be justified on the basis that both DOGs and SMGs are dust obscured populations of galaxies, and may therefore have similar SEDs. Indeed, the median IRAC flux densities of the DOGs and SMGs in GOODS-N are almost identical.
 
We first tested this photometric redshift technique using the Bootes DOGs with spectroscopic redshifts (D08). 
We removed the DOGs which show a power-law in IRAC since this method relies on the presence of the stellar bump to estimate the redshift. This sub-sample of 29 Bootes DOGs has a median spectroscopic redshift of 1.9 (interquartile range 1.6--2.1) and the photometric redshifts using Equation 2 of Pope et al.~(2006) have a distribution with a median of 1.8 (interquartile range 1.4--2.1). 
Comparing the spectroscopic and photometric redshifts for the individual Bootes DOGs we measure a scatter of $\sigma(\Delta z/(1+z))=0.3$ with no obvious biases in the photometric method.
This photometric redshift estimator appears to do a reasonable job in estimating the redshift distribution of DOGs, although on an individual basis the redshifts are still quite uncertain. The redshift distribution for the GOODS-N DOGs using Equation 2 of Pope et al.~(2006) is shown in Fig.~\ref{fig:zdist} (thin solid distribution) along with the subset of DOGs with IRS spectroscopic redshifts (thick solid histogram). The photometric redshift distribution confirms that most of our GOODS-N DOGs sample lies between $z=1.5$--2.5. 

For the rest of this paper, we focus on studying the multi-wavelength properties assuming the average DOG is at $z=2$.

\subsection{AGN and star formation activity}
\label{sec:agnsf}

\begin{figure}
\plotone{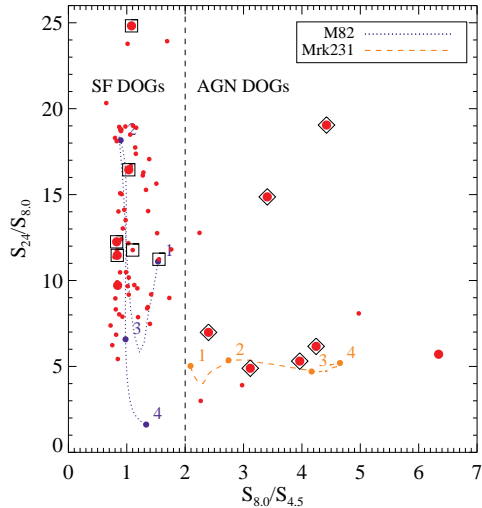}
\caption{{\it Spitzer} color-color diagram used to separate the SF and AGN dominated DOGs. Large circles are the DOGs with $S_{24}>300\,\mu$Jy and small circles are DOGs with $S_{24}=100$--300$\,\mu$Jy. 
Open squares and diamonds are DOGs with IRS spectra classified as SF and AGN dominated, respectively. 
Based on DOGs with IRS spectra we classify SF DOGs as having $S_{8.0}/S_{4.5}<2$ (vertical dashed line); 80\% of DOGs satisfy this criterion. The colors of the M82 (starburst galaxy, Forster Schreiber et al.~2003) and Mrk 231 (AGN-dominated ULIRG, Rigopoulou et al.~1999) as a function of redshift are plotted as the dotted and dashed curves, respectively, with the numbers corresponding to the redshift. 
\label{fig:cc}}
\end{figure}

The IRS spectra also help to quantify the contribution from AGN and star formation (SF) activity to the mid-IR luminosity of DOGs. We use the same spectral decomposition described in Pope et al.~(2008) and classify an object as AGN-dominated if $>50\%$ of the mid-IR luminosity is coming from the continuum component. We find that of the 12 DOG IRS spectra, 6 are AGN dominated, and 6 are SF dominated. 
As was done in Pope et al.~(2008, see also Ivison et al.~2004), we plot these DOGs on a {\it Spitzer} color-color plane (Fig.~\ref{fig:cc}) and find that the IRS SF and AGN dominated sources (open squares and diamonds, respectively) separate very nicely at $S_{8.0}/S_{4.5}=2$. 
This color cut for separating the SF and AGN dominated DOGs is consistent with the simulations of dusty $z=2$ galaxy templates by Sajina et al.~(2005), and also with the `bump' criteria of Farrah et al.~(2008). 

Based on the IRS spectral results, we use this color cut (shown by the dashed line in Fig.~\ref{fig:cc}) to separate the GOODS-N DOGs into SF and AGN dominated classes, and find that 80\% fall into to the SF-dominated class. Interestingly this is the same fraction found using the same diagram for SMGs (Pope et al.~2008). 
The fainter DOGs ($S_{24}=100-300\,\mu$Jy) contain a higher fraction of SF DOGs ($60/66\sim90\%$) than the $S_{24}>300\,\mu$Jy DOGs ($5/13\sim40\%$). 
The median 24$\,\mu$m flux densities for the SF and AGN DOGs are 175 and 310$\,\mu$Jy, respectively.  
This is consistent with the previously noted trend where AGN contribution increases with $S_{24}$ (e.g.~Brand et al.~2006; D08). 

We note that a few of the AGN DOGs have very red $S_{24}/S_{8.0}$ colors inconsistent with the redshifted Mrk 231 spectrum. 
At $z=2$, the 24$\,\mu$m flux can be enhanced by the 7.7$\,\mu$m PAH feature on top of the AGN continuum leading to a higher $S_{24}/S_{8.0}$ color. 
Alternatively, the AGN DOGs could have higher $S_{24}/S_{8.0}$ colors because they contain a heavily obscured AGN; adding more obscuration to a power-law AGN will decrease $S_{8.0}$ more than $S_{24}$ leading to redder colors (Sajina et al.~2007). 
We find that the two AGN DOGs which have both red $S_{24}/S_{8.0}$ colors as well as IRS spectra show a continuum dominated mid-IR spectrum and have no X-ray detection in the deep {\it Chandra} images indicating that they harbor Compton-thick AGN\footnote{One of these sources is also an SMG (source C1, Pope et al.~2008; see also Alexander et al.~2008).}.
This suggests that this {\it Spitzer} color-color plot might also be useful to further separate obscured and unobscured AGN via the $S_{24}/S_{8.0}$ color although much larger samples are needed to test this. 

We can also use the deep X-ray imaging to investigate the presence of X-ray emitting AGN in the DOGs sample. 
We find a higher fraction of X-ray detections in the AGN DOGs than in the SF DOGs (Table \ref{tab:ids}). 
$7/7$ AGN DOGs with X-ray detections have an effective photon index of $\Gamma<1.0$ and an X-ray luminosity of $>10^{42}\,\rm{erg\,s}^{-1}$ (assuming $z\simeq2$) which indicates that the X-rays are coming from obscured AGN emission (Alexander et al.~2005, 2008); only 3/7 SF DOGs with X-ray detections satisfy these criteria. 
Stacking the X-ray undetected DOGs in the central 6.5 arcminute (radius) region of the X-ray image (see Section 4.2 of Alexander et al.~2008 for details) gives a detection in the full and soft bands, but not in the hard band (Table \ref{tab:xray}). 
This implies a hardness ratio (H/S) of $<0.8$ ($3\sigma$ upper limit) which differs from the value of 1.3 found in F08 although their definition of the hard and soft band is slightly different from ours (Table \ref{tab:xray}).
Converting the F08 values to the same bands that we use, their stacking analysis gives a hardness ratio of 0.8 ($\Gamma=1.0$). Our $3\sigma$ upper limit on the hardness ratio of SF DOGs is within the error bar of the measured value of F08. 
A number of factors could lead to a difference in the X-ray stacking results including the area of the X-ray image used in the stacking analysis, the sample size (F08 are stacking 111 objects), and of course the limiting depth at $24\,\mu$m. 
Stacking the mid-IR classified SF and AGN DOGs separately we obtain similar results and conclude that the stacked X-ray hardness ratio of the two sub-samples cannot be distinguished within our uncertainty. 
For the SF DOGs we estimate $L_{0.5-8\,\rm{keV}}=5.7\times10^{41}\,\rm{erg\,s}^{-1}$ and $L_{2-10\,\rm{keV}}=3.9\times10^{41}\,\rm{erg\,s}^{-1}$ from the 1.5--6$\,$keV observed luminosity assuming $\Gamma=1.8$. This X-ray luminosity is several orders of magnitude lower than that of the bright SMGs (Alexander et al.~2005). 
Depending on how much of the X-ray emission is coming from high mass X-ray binaries (HMXBs), we convert the X-ray luminosity of DOGs to SFR and obtain 82--390$\,\rm{M_{\odot}\,yr^{-1}}$ (Bauer et al.~2002 and Persic et al.~2004, respectively). With the small sample of DOGs in GOODS-N, the hard X-rays are poorly constrained and we cannot rule out an X-ray emitting AGN in some DOGs. The high X-ray luminosities of the AGN DOGs are characteristic of AGN activity (e.g.~Alexander et al.~2005). However, contrary to what is implied by F08, most SF DOGs have X-ray emission which is consistent with what is expected from star formation (see Section \ref{sec:sed}). 

\begin{deluxetable}{crrrrr}
\tablecaption{Multi-wavelength detections of DOGs}
\tablewidth{0pt}
\tablehead{
 \colhead{Type} & 
 \colhead{N}  &
 \colhead{1.4$\,$GHz$^{a}$} &  
 \colhead{70$\,\mu$m$^{b}$} &
 \colhead{0.5--8$\,$keV$^{c}$} 
  \label{tab:ids}
}
\startdata
SF DOGs$^{d}$ &  65 & 43  & 0 & 7  \\
AGN DOGs &12  & 11 & 5 & 7 \\
No IRAC$^{e}$ &  2 & 0 & 0 & 0 \\\hline
All DOGs & 79   & 54  & 5 & 14 
\enddata
\tablenotetext{a}{$\sigma\sim5.3\,\mu$Jy (Morrison et al.~in preparation) }
\tablenotetext{b}{$\sigma\sim0.6\,$mJy (Huynh et al.~2007) }
\tablenotetext{c}{$\sigma\sim2.4\times10^{-17}\rm{erg\,cm^{-2}s^{-1}}$ (Table 9 of Alexander et al.~2003) }
\tablenotetext{d}{$S_{8}/S_{4.5}<2$ }
\tablenotetext{e}{These DOGs are outside the uniform IRAC coverage.}
\end{deluxetable}

\begin{deluxetable*}{crrrrrrrrrrrr}
\tabletypesize{\scriptsize}
\tablecaption{X-ray stacking of DOGs}
\tablewidth{0pt}
\tablehead{
 \colhead{Type} & 
 \colhead{N}  &
 \multicolumn{3}{c}{Counts ($10^{-6}\,\rm{s}^{-1}$)$^{a}$} &
    \colhead{H/S} &
 \colhead{$\Gamma$}  &
 \multicolumn{3}{c}{Flux ($10^{-17}\,\rm{cgs})^{b}$} &
    \multicolumn{3}{c}{Luminosity $(10^{42}\,\rm{erg\,s}^{-1})^{c}$} \\
 \colhead{} & 
 \colhead{}  &
 \colhead{0.5--8$\,$keV}   &
 \colhead{0.5--2$\,$keV} &  
 \colhead{2--8$\,$keV} &
 \colhead{} &
 \colhead{} &
 \colhead{0.5--8$\,$keV}   &
 \colhead{0.5--2$\,$keV} &  
 \colhead{2--8$\,$keV} &
 \colhead{1.5--24$\,$keV}   &
  \colhead{1.5--6$\,$keV} &  
 \colhead{6--24$\,$keV} 
 \label{tab:xray}
}
\startdata
All DOGs & 30$^{d}$  & $4.0\,(5.0\sigma)$ &  $2.2\,(5.8\sigma)$ & $<1.7$  & $<0.8$ & $>1.0$ & 4.7 & 1.1 & $<3.9$ & 1.4 & 0.33 & $<1.1$   \\
SF DOGs & 28 & $3.3\,(3.9\sigma)$ &  $2.1\,(5.5\sigma)$ & $<1.8$ & $<0.8$ & $>1.0$   & 3.9 & 1.0 & $<3.9$  & 1.2 & 0.31 & $<1.2$  \\
AGN DOGs & 2  & $12.8\,(3.0\sigma)$ &  $<5.2$ &  $<8.2$ & n/a & n/a  & 15 & $<2.6$ & $<19$ & 4.4 & $<0.77$  & $<5.6$ 
\enddata
\tablenotetext{a}{We list the counts in each band if $>3\sigma$ otherwise we list the $3\sigma$ upper limit. }
\tablenotetext{b}{Assuming $\Gamma=1.4$. }
\tablenotetext{c}{Rest-frame X-ray luminosity assuming $z=2$. }
\tablenotetext{d}{We restrict the stacking analysis to the central 6.5 arcmin region of the Chandra images.}
\end{deluxetable*}

\begin{deluxetable}{crrrc}
\tablecaption{Average flux densities of GOODS-N SF DOGs}
\tablewidth{0pt}
\tablehead{
\colhead{Wavelength ($\mu$m)} & 
\colhead{Flux (mJy)}  
\label{tab:stack}
}
\startdata
24 & 0.17$\pm$0.06  \\
70 &  0.44$\pm$0.11 \\
160 &  6.6$\pm$2.5  \\
850 & 0.95$\pm$0.30 \\
1200 & 0.61$\pm$0.10 
\enddata
\end{deluxetable}

\subsection{IR SED}
\label{sec:sed}
In order to constrain the IR luminosities of DOGs we need data in the far-IR and submm. We have shown that only a handful of DOGs are detected in the far-IR and/or submm, so we perform a stacking analysis to obtain the average SED shape and $L_{\rm{IR}}$ for DOGs. For the following analysis we focus only on the SF DOGs since there are not enough AGN DOGs in GOODS-N to obtain a representative composite SED.

\begin{figure}
\plotone{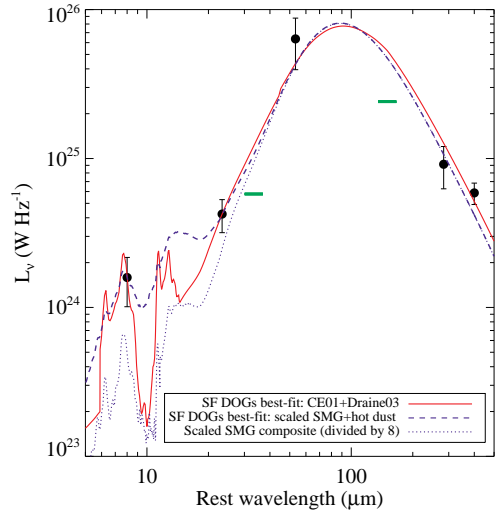}
\caption{Composite SED of SF DOGs: we fit the average fluxes (solid points) of SF DOGs to the CE01+Draine models (solid curve). The dotted curve is a normalized (divided by a factor of 8) composite SED for SMGs (Pope et al.~2008), and the dashed curve is the scaled SMG composite with additional hot ($T=350\,$K) dust. The short horizontal lines indicate the $5\sigma$ depths of the planned deep surveys at 100 and 450$\,\mu$m with Herschel/PACS and SCUBA-2 and show that the majority of DOGs will be detected by these surveys. 
\label{fig:sed}}
\end{figure}

Fig.~\ref{fig:sed} shows a composite SED for the SF DOGs, where we plot the stacked average or median values of $S_{24}$, $S_{70}$, $S_{160}$, $S_{850}$ and $S_{1200}$ (Table \ref{tab:stack})\footnote{Note that not all DOGs are used in the stack at each wavelength, since the samples for stacking were dependent on the coverage and depths of the multi-wavelength maps.} assuming the average DOG is at $z=2$. The solid curve is the best-fit SED to the solid points where we have fit scaled Chary \& Elbaz (2001, CE01) templates with additional extinction from the Draine (2003) models (see Pope et al. 2006 for further details on the SED fitting). 
The dotted curve is a composite SED for bright ($S_{850}>5\,$mJy) SMGs with mid-IR spectra (Pope et al.~2008), scaled down by a factor of 8, which matches the DOGs SED in the far-IR and submm, but is too faint at 24$\,\mu$m and 70$\,\mu$m observed. 
This is consistent with the results of Sajina et al.~(2008) where bright ($S_{24}\gtrsim1\,$mJy) high redshift ULIRGs are rarely detected at (sub)mm wavelengths (see also Lutz et al.~2005). 
The excess emission in the mid-IR relative to SMGs in the sub-sample of strong PAH sources from Sajina et al.~(2008) accounts for 30\% of the total IR luminosity. 
If we add a hot ($T=350\,$K) dust component to the SMG composite then we obtain a good fit to the SF DOGs (dashed curve). 
The additional hot dust component accounts for less than 10\% of the total IR luminosity and could be due to SF or AGN activity (Tran et al.~2001). With IRS spectra existing for only a small subset of the brightest sources in our sample we cannot say whether this mid-IR excess is due to hot dust or enhanced PAH emission (the SNR of the IRS spectra is not high enough to differentiate the PAH equivalent widths between the SMGs and the DOGs). However, regardless of the source of the excess, our best-fit SED shows that the total IR luminosity in these objects is dominated by the cold dust component, presumably fueled by star formation. 
The 8--1000$\,\mu$m total $L_{\rm{IR}}$ for both the solid curve and the dashed curve is $1\times10^{12}\,\rm{L_{\odot}}$ and the average dust temperature is $\sim32\,$K. The IR luminosity implied by the median radio flux of DOGs ($\sim20\,\mu$Jy) and the radio-IR correlation is consistent with this estimate. 

In order to estimate the uncertainty in $L_{\rm{IR}}$ for the SF DOGs we perform 1000 Monte Carlo simulations where we randomly sample each data point assuming a Gaussian with mean equal to the stacked value and $\sigma$ equal to the uncertainty in this value. We also include the redshift uncertainty assuming a Gaussian redshift distribution centered on 2 with $\sigma=0.3$ (Fig.~\ref{fig:zdist}). The resulting distribution of $L_{\rm{IR}}$ is $(1.1\pm0.5)\times10^{12}\,\rm{L_{\odot}}$, where the error is the $1\sigma$ uncertainty\footnote{If we assume $\sigma=0.5$ for a Gaussian redshift distribution centered on 2 then we get $L_{\rm{IR}}=(1.1\pm0.7)\times10^{12}\,\rm{L_{\odot}}$.}.
This $L_{\rm{IR}}$ for the SF DOGs implies a star formation rate of $200\,\rm{M_{\odot}\,yr^{-1}}$, using the Kennicutt (1998) relation. 
This is consistent with the range of estimates from the X-ray emission given that X-ray SFR relation depends strongly on the relative contributions from HMXBs and LMXBs to the X-ray emission (e.g.~Persic et al.~2004). 

The short horizontal lines in Fig.~\ref{fig:sed} indicate the $5\sigma$ limits of the deepest surveys to be done with {\it Herschel}/PACS at 100$\,\mu$m (Pilbratt 2001) and JCMT/SCUBA-2 at 450$\,\mu$m (Holland et al.~2006); the majority of DOGs will be detected. 
These surveys will put constraints on the infrared luminosities and dust temperatures of individual galaxies without the need for stacking.

\section{Discussion}

The average $L_{\rm{IR}}$ derived for the DOGs using submm and far-IR measurements is a factor of $\sim4$ times smaller than that calculated in D08 using a conversion from $S_{24}$ observed (aka $L_{8}$ rest $=\nu L_{\nu}|_{8\,\mu m}$) to $L_{\rm{IR}}$. Part of this is because our average $L_{8}$ is lower, since we push three times deeper at $24\,\mu$m, and part is because of the assumed conversion between mid-IR and total IR luminosity. Based on our best-fit SED, we calculate that $L_{\rm{IR}}/L_{8}\simeq7$ (quartile range from Monte Carlo simulations is 5--10) which is within the lower range of conversion factors assumed in D08 ($L_{\rm{IR}}/L_{8}$=5--15). On the other hand the SMGs from Pope et al.~(2008) with $S_{850}>5\,$mJy have an average $L_{\rm{IR}}/L_{8}$ of $\sim20$. The conversion between 24$\,\mu$m flux and $L_{\rm{IR}}$ is uncertain for high redshift galaxies since it relies on local galaxy templates. Observations of high redshift ULIRGs indicate an evolution in SED shapes from local ULIRGs (e.g.~Pope et al.~2006, 2008; Rigby et al.~2008). The overestimate of $L_{\rm{IR}}$ (and SFR) using only $S_{24}$ has been noted for bright high redshift ULIRGs (e.g.~Papovich et al.~2007, Daddi et al.~2007). It is clear from our full SED fits for DOGs and SMGs that a uniform conversion cannot be applied to all high redshift ULIRGs and there must be additional parameters, other than mid-IR luminosity, which are needed to determine the total $L_{\rm{IR}}$. Progress in determining these parameters will be facilitated with future wide-field far-IR and submm surveys with {\it Herschel Space Observatory} and SCUBA-2, for example. 

As discussed in D08, DOGs with $S_{24}>300\,\mu$m have similar surface densities to SMGs with $S_{850}>6\,$mJy (Coppin et al.~2006). The number density of DOGs in GOODS-N down to $S_{24}=100\,\mu$Jy is 0.5$\,$arcmin$^{-2}$, 6 times more than the shallower sample in D08. Given the average SFR from $L_{\rm{IR}}$ for the SF DOGs ($200\,\rm{M_{\odot}\,yr^{-1}}$), we calculate a SFRD of $=0.01\,\rm{M_{\odot}yr^{-1}Mpc^{-3}}$ at $z=2$. Depending on the value adopted for the total SFRD at $z=2$ (e.g.~Chary \& Elbaz 2001; Caputi et al.~2007), SF DOGs contribute 5--10\% of the total SFRD at $z=2$. This result appears to conflict with D08 who suggest that ($S_{24}>300\,\mu$Jy) DOGs contribute 25\% of the total IR luminosity density at $z=2$. However, if we remove the AGN DOGs from the D08 sample and use the lower conversion factor of $L_{\rm{IR}}/L_{8}\simeq7$, this 25\% becomes 7\% which is consistent. The AGN DOGs will further contribute to the SFRD, however the small number of AGN DOGs in GOODS-N does not allow us to put constraints on their average $L_{\rm{IR}}$ and SFR. While the simple DOGs selection can be used in many of the deep {\it Spitzer} surveys to isolate large samples of high redshift ULIRGs, this selection alone does not resolve the bulk of the SFRD at $z=2$. For comparison, bright ($S_{850}>5\,$mJy) SMGs contribute $0.02\,\rm{M_{\odot}yr^{-1}Mpc^{-3}}$, with fainter ($S_{850}>2\,$mJy) SMGs contributing $0.05\,\rm{M_{\odot}yr^{-1}Mpc^{-3}}$ at $z=2$ (Wall et al.~2008).
In addition to the DOG samples, several different selection criteria using {\it Spitzer} data have been presented in the literature to isolate ULIRGs at high redshifts (e.g.~Yan et al.~2007; Farrah et al.~2008), although none of these select a population as numerous as the DOGs. 
Table \ref{tab:ids} shows that 70\% of DOGs are detected in the radio. Since DOGs by definition are faint in the optical, these radio-detected DOGs would be similar to the optically-faint radio galaxies (OFRGs) discussed in Chapman et al.~(2002, 2004). 

SMGs are thought to be galaxies in the early stage of a massive merger (e.g., Conselice et al.~2003; Pope et al.~2008; Tacconi et al.~2008), and D08 propose that bright DOGs might be a later stage in the merger. In support of this, we find that $\sim30\%$ of SMGs meet the DOGs criteria and all 3 SMGs which have $>50\%$ AGN contribution in the mid-IR (Pope et al.~2008) are in our sample of AGN DOGs. This implies that DOGs selection preferentially picks up the more AGN dominated SMGs, although these are among the most luminous DOGs. 
The average SF DOG shows additional mid-IR emission compared to the normalized SMG SED, which may be enhanced PAH emission or hot dust heated by an AGN or star formation. 
Regardless of the source of the mid-IR excess emission (which accounts for $<10\%$ of the total IR luminosity), the average $L_{\rm{IR}}$ and X-ray luminosity of the SF DOGs is several times less than that of most SMGs indicating that the average DOG is not likely to evolve from SMGs. 
Fig.~\ref{fig:bzk} shows that most DOGs satisfy the {\it BzK} selection; while they have ULIRG-like luminosities {\it BzK} galaxies are thought to be forming stars continuously over longer timescales and do not necessarily require a major merger as as catalyst for star formation (Daddi et al.~2008).
In summary, we remind the reader that this analysis is focussed on the average properties of DOGs in GOODS-N; while the average DOG is less luminous than $S_{850}>5\,$mJy SMGs, some fraction of the DOGs are related to SMGs as shown in the 30\% of SMGs which meet the DOGs criteria. 
In order to obtain a submm sample of comparable number density to the $S_{24}>100\,\mu$Jy DOGs sample requires a survey down to $S_{850}>3\,$mJy (Coppin et al.~2006). This will be achieved with future deep SCUBA-2 surveys and allow for a more detailed comparison between DOGs and submm-emitting galaxies.

\section{Summary}

From a sample of 79 faint ($S_{24}>100\,\mu$Jy) DOGs in GOODS-N ($0.5\,$arcmin$^{-2}$), we find that almost all satisfy the criteria for Compton-thick AGN from F08. However, based on {\it Spitzer} spectroscopy and photometry, we show that 80\% of are likely dominated by star formation. The stacked X-ray emission from the mid-IR classified star forming DOGs is consistent with what is expected from star formation. 

The IRS spectra and {\it Spitzer} photometric redshifts confirm that these faint DOGs lie in a tight redshift distribution around $z\sim2$. Stacking the mid-IR, far-IR and submm flux of the star forming DOGs, we derive an average SED with $L_{\rm{IR}}\sim1\times10^{12}\,\rm{L_{\odot}}$, 8 times less luminous than most bright ($S_{850}>5\,$mJy) SMGs.
The composite SED of DOGs has a similar shape to that of SMGs in the far-IR (dust temperature of around 30$\,$K) but has a higher mid-IR to far-IR flux ratio ($L_{\rm{IR}}/L_{8}\simeq7$ compared to $L_{\rm{IR}}/L_{8}\simeq20$ for SMGs). This suggests that there is a wide range of $L_{\rm{IR}}/L_{8}$ conversions in $z=2$ galaxies which need to be considered when interpreting the total IR luminosity density and SFRD from 24$\,\mu$m surveys. 

The average star forming DOG has a star formation rate of $200\,\rm{M_{\odot}\,yr^{-1}}$ which amounts to a contribution of $0.01\,\rm{M_{\odot}\,yr^{-1}Mpc^{-3}}$ (or 5--10\%) to the star formation rate density at $z\sim2$. 

This paper has relied strongly on stacking analysis to obtain average properties of DOGs (Fig.~\ref{fig:sed}). Future deep surveys with {\it Herschel} and SCUBA-2 will detect the majority of DOGs putting constraints on their individual IR luminosities and dust temperatures. 

\acknowledgments
We thank the referee for constructive comments on this paper. 
We are grateful to Emanuele Daddi and Anna Sajina for insightful discussions. 
AP acknowledges support provided by NASA through the {\it Spitzer Space Telescope} Fellowship Program, through a contract issued by the Jet Propulsion Laboratory, California Institute of Technology under a contract with NASA. The research of AP, AD, MB, MED is supported in part by NOAO, which is operated by the Association of Universities for Research in Astronomy (AURA) under a cooperative agreement with the National Science Foundation. DMA thanks the Royal Society for support. 
This work is based on observations made with the {\it Spitzer Space Telescope}, which is operated by the Jet Propulsion Laboratory, California Institute of Technology under a contract with NASA. Support for this work was provided by NASA through an award issued by JPL/Caltech.
We acknowledge L.~Simard for leading the GOODS-N WIRCAM $K$-band proposal and L.~Albert and the Terapix team for help on the WIRCAM data processing.


\end{document}